\documentclass[aps, reprint]{revtex4-2}

\usepackage{amsmath}
\usepackage{graphicx} 
\usepackage[colorlinks,linkcolor=red,anchorcolor=blue,citecolor=blue]{hyperref}
\usepackage[title,titletoc]{appendix}
\usepackage{balance}
\usepackage{mathtools}
\usepackage{bbm}
\usepackage{bm}
\usepackage{subfigure}
\usepackage{hyperref}
\newtheorem{theorem}{Theorem}

\begin{document}

\title{A Gell-Mann \& Low Theorem Perspective on Quantum Computing: New Paradigm for Designing Quantum Algorithm}

\author{Chun-Tse Li$^{1,2,3}$}
\thanks{These authors contributed equally to this work}
\author{T. Tzen Ong$^{2}$,$^{*}$}
\email{tzenong@gmail.com}
\author{Lucas Wang$^4$}
\author{Ming-Chien Hsu$^1$}
\author{Hsin Lin$^{2}$}
\email{nilnish@gmail.com}
\author{Min-Hsiu Hsieh$^1$}

\affiliation{$^1$Hon Hai Quantum Computing Research Center, Taipei, Taiwan}
\affiliation{$^2$Institute of Physics, Academia Sinica, Taipei 115201, Taiwan}
\affiliation{$^3$Department of Electrical Engineering,
National Taiwan University,
Taipei 106, Taiwan}
\affiliation{$^4$Pritzker School of Molecular Engineering, University of Chicago, Chicago, IL 60637, USA}


\begin{abstract}
The Gell-Mann \& Low theorem is a cornerstone of Quantum Field Theory (QFT) and condensed matter physics, and many-body perturbation theory is a foundational tool for treating interactions. However, their integration into quantum algorithms remains a largely unexplored area of research, with current quantum simulation algorithms predominantly operating in the Schrödinger picture, leaving the potential of the interaction picture largely untapped. Our Variational Interaction-Picture S-matrix Ansatz (VIPSA) now fills this gap, specifically in the context of the Fermi-Hubbard model —a canonical paradigm in condensed matter physics which is intricately connected to phenomena such as high-temperature superconductivity and Mott insulator transitions. 

This work offers a new conceptual perspective for variational quantum computing based upon the Gell-Mann \& Low theorem. We achieve this by employing an innovative mathematical technique to explicitly unfold the normalized S-matrix, thereby enabling the systematic reconstruction of the Dyson series on a quantum computer, order by order. This method stands in contrast to the conventional reliance on Trotter expansion for adiabatic time evolution, marking a conceptual shift towards more sophisticated quantum algorithmic design. We leverage the strengths of the recently developed ADAPT-VQE algorithm, tailoring it to reconstruct perturbative terms effectively. Our simulations indicate that this method not only successfully recovers the Dyson series but also exhibits robust and stable convergence. We believe that our approach shows great promise in generalizing to more complex scenarios without increasing algorithmic complexity. 
\end{abstract}

\maketitle

\section{Introduction}

Understanding and addressing the complexities of strongly interacting quantum systems is a pursuit of both fundamental scientific importance and potential technological innovation. These systems, marked by intense particle interactions, serve as a breeding ground for a range of complex emergent phenomena and the discovery of novel states of matter. Notable among these are non-Abelian topological states~\cite{moore1991nonabelions, NayakFreedmanRMP2008}, spin liquids~\cite{anderson1973resonating, balents2010spin, savary2016quantum}, and high-temperature superconductors~\cite{bednorz1986possible, lee2006doping, wu1987superconductivity}, each offering a unique glimpse into the intricate workings of quantum mechanics. Additionally, the methodologies and algorithms honed for these strongly correlated systems find broader applicability, extending to various combinatorial problems~\cite{lucas2014ising, farhi2014quantum, wecker2016training, farhi2017quantum}, thereby bridging the gap between theoretical physics and practical computational solutions.

Elucidating the fundamental physics underlying these systems is a formidable challenge primarily due to the exponential growth in dimensionality coupled with the strength of electron interactions, which significantly hampers the effectiveness of classical numerical methods~\cite{KotliarGeorgesPRB1992, white1992density, georges1996dynamical, kotliar2006electronic, schollwock2005density, foulkes2001quantum}. In this complex and intricate scenario, quantum computing, with its inherent ability to manipulate operations within an exponentially large Hilbert space, naturally emerges as a powerful ally.

Current fervent research efforts in quantum computing have catalyzed the swift evolution of quantum algorithms. Numerous methodologies have been proposed to simulate quantum dynamics, with seminal works~\cite{childs2012hamiltonian, berry2015hamiltonian, berry2015simulating, low2017optimal, low2019hamiltonian, gilyen2019quantum}. Despite significant advancements, a key challenge persists in the resilience of quantum algorithms against noise. The prevalent Noisy Intermediate-Scale Quantum (NISQ) \cite{preskill2018quantum, bharti2022noisy} computers exhibit noise levels that impede the effective deployment of these algorithms, highlighting a crucial gap in their practical application under real-world conditions.

In the Noisy Intermediate-Scale Quantum (NISQ) era, Variational Quantum Algorithms (VQAs)~\cite{peruzzo2014variational, tilly2022variational} have emerged as a leading strategy for quantum computing, particularly with the development of methods like the Quantum Approximation Optimization Algorithm (QAOA)~\cite{farhi2014quantum, zhou2020quantum, hadfield2019quantum} and the Hamiltonian Variational Ansatz (HVA)~\cite{wecker2015progress}. These approaches focus on adiabatic time evolution and efficient identification of the Hubbard Hamiltonian's ground state, with subsequent studies further refining these techniques~\cite{cade2020strategies, wiersema2020exploring, stanisic2022observing, vogt2020preparing, choquette2021quantum}. Additionally, several algorithms have been proposed to address the Fermi-Hubbard model on quantum computers, focusing on reducing circuit complexity for preparing the non-interacting ground state and implementing basis transformation, as highlighted in \cite{babbush2018low, wecker2015solving, jiang2018quantum}.
In parallel, the adaptive structure ansatz, particularly in simulating electronic structure problems in quantum chemistry, has shown significant advancements. The ``Fermionic ADAPT-VQE" approach by Grimsley et al.~\cite{grimsley2019adaptive} innovatively constructs the ansatz circuit by selecting high-impact operators from a pool, mainly derived from single and double excitation cluster operators within Unitary Coupled Cluster (UCC) theory~\cite{taube2006new, romero2018strategies, lee2018generalized}. This method has led to compact ansatz circuits with fewer parameters, potentially circumventing the barren plateau problem~\cite{mcclean2018barren, holmes2022connecting}. Recent studies~\cite{tang2021qubit, yordanov2021qubit} have suggested adapting this method to the Pauli representation, which may offer even more compact ansatzes at the cost of additional parameters and possibly provides less physical intuition when selecting operators. Despite most research using the Schrödinger picture, notable exceptions like ~\cite{ryabinkin2020iterative} have explored iterative methods for Hamiltonian transformation in the Heisenberg picture, yet discussions on simulations within the interaction picture~\cite{low2018hamiltonian} and integration of perturbation theory in quantum simulations remain limited.

In contrast, we propose a conceptually new perspective for designing VQAs that is based upon the (Dirac) interaction-picture. At the heart of our methodology lies the Gell-Mann \& Low theorem~\cite{gell1951bound, hepp1969theorie, nenciu1989adiabatic} — a fundamental tool in Quantum Field Theory (QFT) that delineates an exact mapping from non-interacting to interacting eigenstates. Despite its seminal importance in QFT, its application within quantum computing remains an untapped potential. We address this by formulating a novel operator-expansion of the S-matrix within the Gell-Mann \& Low theorem, allowing us to side-step the standard Wick's theorem and Feynman diagram methodology, and this enable us to asymptotically approach the complete Dyson series using a quantum computer. Inspired by the S-matrix, we propose a functional form for our VIPSA ansatz, formed by exponentiating the generators derived from the interaction Hamiltonian. Through the theoretical analysis, we deduce that the Dyson series is actually order by order recoverable by applying a sequence of ansatz unitaries. However, the exact gate sequences and parameter assignment required is combinatorically challenging to solve, and instead, we implement a variational version of the VIPSA by relying on the recently developed ADAPT-VQE to iteratively choose the variationally optimal gate sequence for the S-matrix. Our focus is on the Fermi-Hubbard model \cite{hubbard1964electron, arovas2022hubbard, esslinger2010fermi}, a model that epitomizes strong correlations within condensed matter theory.  Successfully applying our methods to this model highlights the adaptability and effectiveness of our approach. These results, encompassing both theoretical innovation and algorithmic practicality, stand as the key contributions of our work, potentially paving new pathways in the understanding and simulation of complex quantum systems.

\section{Preliminary}
In this section, we delve into the fundamental concepts of the interaction picture and the Gell-Mann and Low theorem, which are critical in the realms of quantum field theory and condensed matter physics. For detail information, there are some comprehensive references \cite{gell1951bound, bruus2004many, kleinert2016particles}.

\subsection{Introduction to the Interaction Picture}
The interaction picture, also known as the Dirac picture, plays a crucial role in quantum mechanics, particularly in the context of quantum field theory and many-body physics. It represents a hybrid approach, blending aspects of the Schrödinger and Heisenberg pictures, and is particularly useful for dealing with systems involving interactions, as it simplifies the treatment of time evolution. In this picture, we describe a system using a time-independent non-interacting Hamiltonian, $H_0$, alongside a time-dependent interacting Hamiltonian, $H_1(t)$. This setup enables the explicit calculation of $H_0$'s eigenstates:
\begin{equation}
    H = H_0 + H_1(t),\hspace{1cm} H_0|n\rangle = E_0|n\rangle.
\end{equation}
The essence of the interaction picture lies in its division of time evolution into two components: one governed by $H_0$ and the other by $H_1(t)$. In this framework, operators evolve under the influence of $H_0$, while state vectors follow the dynamics introduced by $H_1(t)$:
\begin{equation}
    \left\{
    \begin{array}{l}
         |\Psi_\mathcal{I}(t)\rangle \equiv e^{iH_0t}|\Psi_\mathcal{S}(t)\rangle, \\
         \\
         O_{\mathcal{I}}(t)\equiv e^{iH_0t}O_{\mathcal{S}}e^{-iH_0t}. 
    \end{array}
    \right.
\end{equation}
where the subscript $\mathcal{I}$ and $\mathcal{S}$ represent the state or operator in interaction picture and Schrodinger picture respectively. 

A state vector in the interaction picture follows this equation of motion:
\begin{align}
\label{eq:state EOM}
    i\partial_t |\Psi_\mathcal{I}(t)\rangle &= i\partial_t e^{iH_0t}|\Psi_\mathcal{S}(t)\rangle 
    \nonumber \\
    &=e^{iH_0t}(-H_0+H)|\Psi_\mathcal{S}(t)\rangle
    \nonumber \\
    &=e^{iH_0t}H_1e^{-iH_0t}e^{iH_0t}|\Psi_\mathcal{S}(t)\rangle
    \nonumber \\
    &=H_{1,\mathcal{I}}(t)|\Psi_\mathcal{I}(t)\rangle.
\end{align}
The time-evolution unitary in the interaction picture is derived as:
\begin{align}
    &|\Psi_\mathcal{I}(t)\rangle=U_\mathcal{I}(t,t_0)|\Psi_\mathcal{I}(t_0)\rangle 
    \nonumber \\
    &\Rightarrow U_{\mathcal{I}}(t,t_0)=e^{iH_0t}e^{-iH(t-t_0)}e^{-iH_0t}.
\end{align}
Using the relation in Eq.~(\ref{eq:state EOM}), the operator $U_{\mathcal{I}}(t,t_0)$ satisfies the following equation of motion:
\begin{align}
\label{eq: unitary EOM}
    i\partial_t U_{\mathcal{I}}(t,t_0)=H_{1,\mathcal{I}}(t)U_{\mathcal{I}}(t,t_0), \hspace{0.5cm} U_{\mathcal{I}}(t_0, t_0)=\mathbbm{1}.
\end{align}
By integrating the differential equation in Eq.~(\ref{eq: unitary EOM}), we get the integration equation:
\begin{align}
    U_{\mathcal{I}}(t,t_0)=\mathbbm{1}-i\int_{t_0}^{t}dt' H_{1,\mathcal{I}}(t')U_{\mathcal{I}}(t',t_0).
\end{align}
which we can solve recursively:
\begin{align}
\label{eq: Perturbation 1}
    U_{\mathcal{I}}(t,t_0)=&\mathbbm{1}-i\int_{t_0}^{t}dt_1H_{1,\mathcal{I}}(t_1)+
    \nonumber \\
    &(-i)^2\int_{t_0}^{t}dt_1\int_{t_0}^{t_1}dt_2 H_{1,\mathcal{I}}(t_2)H_{1,\mathcal{I}}(t_1)+\cdots.
\end{align}
By relabelling the dummy variable $t_1\cdots t_n$, we can see that the second order perturbation can be written into a more compact form:
\begin{align}
    \frac{(-i)^2}{2!}\int_{t_0}^{t}dt_1\int_{t_0}^{t}dt_2 \mathcal{T}\biggl(H_{1,\mathcal{I}}(t_2)H_{1,\mathcal{I}}(t_1)\biggr)+\cdots.
\end{align}
where we introduce the time ordered operator $\mathcal{T}$:
\begin{equation}
    \mathcal{T}\biggl(B(t_2)A(t_1)\biggr) =
    \left\{
    \begin{array}{cc}
         B(t_2)A(t_1)  &  \hspace{0.5cm} \text{if } t_2 > t_1, \\ 
         \\
         A(t_1)B(t_2)  &  \hspace{0.5cm} \text{if } t_2 < t_1.
    \end{array}
    \right.
\end{equation}
Therefore the perturbation expansion can be written in the notation of time ordered integral:
\begin{footnotesize}
    \begin{align}
    \label{eq:pertubation expansion of time evolution operator}
    U_{\mathcal{I}}(t, t_0) &= \sum_{n=0}^{\infty} \frac{(-i)^n}{n!}\int_{t_0}^{t}dt_n\cdots\int_{t_0}^{t}dt_1 \mathcal{T}\biggl(H_{1,\mathcal{I}}(t_n)\cdots H_{1,\mathcal{I}}(t_1)\biggr)
    \nonumber \\
    &=\mathcal{T}\biggl(e^{-i\int_{t_0}^tdt'H_{1,\mathcal{I}}(t')}\biggr).
\end{align}
\end{footnotesize}
The main use of the time-evolution operator lies in its application to scattering processes. Furthermore $U_{\mathcal{I}}(t, t_0)$ can be employed to calculate the energy
shift of a bound level under the influence of an interaction. This will be discussed in the Gell-Mann \& Low theorem in the later section.

\subsection{Introduction to S-matrix}
The S-matrix, or scattering matrix, is a cornerstone concept in quantum field theory and particle physics. It describes the probability amplitude for a processe that evolves from the initial state to final state under the influence of interaction, specially in scenarios like particle scattering experiments. The essence of the S-matrix lies in its ability to connect the state of a system at one point in time to its state at another point. 

In a typical scattering experiment, we are interested in the evolution of particles' state vectors. Before the collision or interaction occurs, the system is in a free initial state $|\Phi_i\rangle$, represented by a state vector $|\Psi_\mathcal{I}(t)\rangle$ as $t\to -\infty$. 
\begin{align}
    \lim_{t\to-\infty}|\Psi_{\mathcal{I}}(t)\rangle=|\Phi_i\rangle.
\end{align}
The S-matrix element is defined by calculating the projection of the output state vector $|\Psi_{\mathcal{I}}(t)\rangle$ as $t\to+\infty$ to the final state $|\Phi_f\rangle$ , labelled with the quantum number $f$: 
\begin{align}
    S_{fi}&=\lim_{t\to+\infty}\langle\Phi_f|\Psi_\mathcal{I}(t)\rangle 
    \nonumber\\
    &=\langle\Phi_f|S|\Phi_i\rangle
    \nonumber\\
    &=\lim_{t_2\to+\infty}\lim_{t_1\to-\infty}\langle\Phi_f|U_\mathcal{I}(t_2,t_1)|\Phi_i\rangle,
\end{align}
where the S-operator is the time evolution operator that evolves from $t=-\infty$ to $t=\infty$:
\begin{equation}
    S=U_{\mathcal{I}}(+\infty, -\infty).
\end{equation}
We can formally derive the perturbative expansion of S-operator from the Eq.~(\ref{eq:pertubation expansion of time evolution operator}):
\begin{align}
    S = \sum_{n=0}^{\infty} \frac{(-i)^n}{n!}\int_{-\infty}^{\infty}dt_n\cdots\int_{-\infty}^{\infty}dt_1 \mathcal{T}\biggl(H_{1,\mathcal{I}}(t_n)\cdots H_{1,\mathcal{I}}(t_1)\biggr).
\end{align}
Physically, the S-matrix has profound implications. Its elements correspond to the probability amplitudes of transitioning from specific initial states to specific final states in a scattering process. The unitarity of the S-matrix, a fundamental property indicating that the matrix is its own inverse, ensures that probability is conserved in these processes. This unitarity is deeply intertwined with fundamental conservation laws in physics.

\subsection{Gell-Mann \& Low Theorem}
The Gell-Mann and Low theorem represents a cornerstone in quantum field theory, This theorem is 
essential for understanding how quantum systems evolve with the introduction of an interaction.

Consider a quantum system initially described by a non-interacting Hamiltonian \( H_0 \). The introduction of an interaction, described by \( H_1 \), transforms 
the system's Hamiltonian to \( H = H_0 + H_1 \). The theorem explores how an 
eigenstate \( |\Phi_0\rangle \) of \( H_0 \) evolves into an eigenstate \( |\psi\rangle \) of the full Hamiltonian \( H \), as given by the Schrödinger equation:
\begin{equation}
    (H_0 + H_1)|\psi\rangle = E|\psi\rangle.  
\end{equation}
Correspondingly, the unperturbed state \( |\Phi_0\rangle \) satisfies:
\begin{equation}
   H_0|\Phi_0\rangle = E_0|\Phi_0\rangle.  
\end{equation}

To model the transition from a non-interacting state to an interacting one, a 
coupling constant \( g \) is introduced, facilitating a smooth evolution from 
\( g = 0 \) (non-interacting) to \( g = 1 \) (fully interacting). The theorem utilizes 
S-matrix to map this transition, particularly focusing on the adiabatic 
limit (\( \eta \to 0 \)) to ensure a smooth evolution.
The normalized S-matrix, in the context of the Gell-Mann \& Low theorem, transforms the non-interacting ground state \( |\Phi_0\rangle \) into the interacting 
ground state \( |\psi^{(\pm)}\rangle \):
\begin{equation}
\label{eq:GML theorem}
    |\psi^{(\pm)}\rangle = \lim_{\eta \to 0} \frac{U_\mathcal{I}^\eta(0, \mp\infty) |\Phi_0\rangle}{z_\eta^{(\pm)}}.  
\end{equation}
Here, \( z_\eta^{(\pm)} \), the normalization factor, is crucial for cancelling out the infinities resulted from S-matrix:
\begin{equation}
    z_\eta^{(\pm)} = \langle \Phi_0 | U_{\mathcal{I}}^\eta(0, \mp\infty) | \Phi_0 \rangle.
\end{equation}

The role of \( z_\eta^{(\pm)} \) extends beyond mere normalization; it also introduces 
an energy shift to the system. This energy shift, a critical aspect of the interacting 
dynamics, is expressed as:
\begin{equation}
    \Delta E = \lim_{\eta \to 0} (\mp i\eta g \frac{\partial}{\partial g} \log z_\eta^{(\pm)}).
\end{equation}
In essence, the Gell-Mann \& Low theorem provides a mathematically rigorous method 
to track the evolution of quantum states in the presence of interactions. By employing 
time evolution operator in the adiabatic limit, it offers a profound insight into the energy shifts and state transformations inherent in quantum field theory.

\begin{figure*}[htbp]
\centering
    \includegraphics[width=\textwidth]{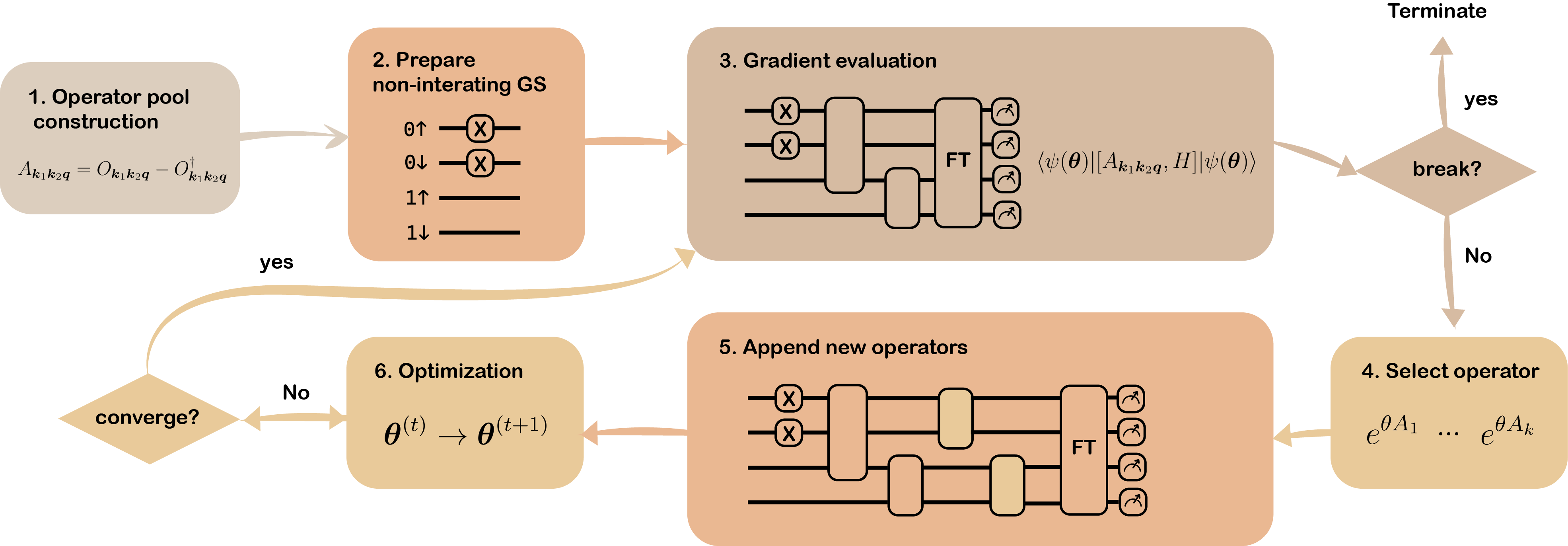}
    \caption{Schematic of the Variational Interaction Picture S-matrix Ansatz. The process begins with the construction of an operator pool (1), followed by the preparation of a non-interacting ground state in k-space (2). Gradient evaluation (3) guides the selection of operators (4) based on the gradient threshold. New operators are then appended to the variational form (5), and the entire ansatz is optimized (6) until convergence is achieved, completing the iterative process. If convergence is met, the algorithm terminates; otherwise, it loops back for further optimization.}
    \label{fig: flow chart}
\end{figure*}

\section{Main Result}
In this section, we provide a succinct overview of our principal theoretical findings and delineate the overarching methodology of the adaptive algorithm. Unlike standard quantum field theory treatments that apply Wick's theorem to obtain expansions in terms of Green's functions, we instead carry out a unique analysis of the time-ordered S-matrix in terms of time-ordered integrals. The causatum of this is an operator expansion of the S-matrix, which is one of our key theoretical contributions. A comprehensive proof of the theoretical results is available in the Appendix for interested readers. 

Our discussion is framed within the momentum space ($k$-space), necessitating the transformation of the Hubbard Hamiltonian into $k$-space representation:
\begin{align}
\label{eq:Hubbard H k-space}
H=\sum_{\bm{k}\sigma}\epsilon_{\bm{k}}c^\dagger_{\bm{k}\sigma}c_{\bm{k}\sigma}+\frac{U}{N}\sum_{\bm{k}_1\bm{k}_2\bm{q}} c^\dagger_{\bm{k}_1+\bm{q}\uparrow}c^\dagger_{\bm{k}_2-\bm{q}\downarrow}c_{\bm{k}_2\downarrow}c_{\bm{k}_1\uparrow}.
\end{align}
Here, $\epsilon_{\bm{k}}$ denotes the dispersion relation in $k$-space which simplifies to $\epsilon_{\bm{k}} = -2t(\cos{k_x}+\cos{k_y})$ under periodic boundary conditions. With the $k$-space representation of the Hubbard Hamiltonian, we can describe our main theorem:
\\
\\
\begin{theorem}
    Given the Gell-Mann \& Low formula as in Eq.~(\ref{eq:GML theorem}), a closed-form expression can be derived by substituting the Hubbard Hamiltonian in Eq.~(\ref{eq: hubbard model real space}):
\begin{align}
    \label{eq:main theorem formula}
    |\psi^{(+)}\rangle &= \lim_{\eta \to 0} \frac{U_\mathcal{I}^\eta(0, -\infty) |\Phi_0\rangle}{\langle \Phi_0 | U_\mathcal{I}^\eta(0, -\infty) | \Phi_0 \rangle}  
    \\
    &=\Tilde{U}_\mathcal{I}(0,-\infty)|\Phi_0\rangle, \nonumber
\end{align}
where we introduce the normalized time evolution operator $\Tilde{U}_{\mathcal{I}}(0,-\infty)$:
\begin{footnotesize}
    \begin{align}
    &\Tilde{U}_{\mathcal{I}}(0,-\infty)\equiv \mathbbm{1}-g\sideset{}{'}\sum_{\bm{k}'_1\bm{k}'_2\bm{q}'}\frac{1}{\epsilon_{\bm{k}'_1\bm{k}'_2\bm{q}'}}O_{\bm{k}'_1\bm{k}'_2\bm{q}'} 
    \\
    &\hspace{2cm}+g^2\sideset{}{'}\sum_{\bm{k}''_1\bm{k}''_2\bm{q}''}\sideset{}{'}\sum_{\bm{k}'_1\bm{k}'_2\bm{q}'}\frac{1}{\epsilon_{\bm{k}'_1\bm{k}'_2\bm{q}'}+\epsilon_{\bm{k}''_1\bm{k}''_2\bm{q}''}}\frac{1}{\epsilon_{\bm{k}'_1\bm{k}'_2\bm{q}'}}
    \nonumber\\
    &\hspace{4.5cm}\times O_{\bm{k}''_1\bm{k}''_2\bm{q}''}O_{\bm{k}'_1\bm{k}'_2\bm{q}'}-\cdots
    \nonumber\\
    &=\sum_{n=0}^{\infty}(-g)^n\sideset{}{'}\sum_{\bm{k}^{(n)}\bm{k}^{(n)}\bm{q}^{(n)}}\cdots\sideset{}{'}\sum_{\bm{k}'_1\bm{k}'_2\bm{q}'}O_{\bm{k}^{(n)}\bm{k}^{(n)}\bm{q}^{(n)}}\cdots O_{\bm{k}'_1\bm{k}'_2\bm{q}'}
    \nonumber
    \nonumber \\
    &\hspace{4cm}\times\prod_{m=1}^{n}\frac{1}{\sum_{j=1}^{m}\epsilon_{\bm{k}^{(j)}_1\bm{k}^{(j)}_2\bm{q}^{(j)}}},
    \nonumber
\end{align}
\end{footnotesize} \\
where $g$ is defined as $\frac{U}{N}$, related to the Hubbard interaction strength and the total number of sites, and $\epsilon_{\bm{k}_1\bm{k}_2\bm{q}}$ and $O_{\bm{k}_1\bm{k}_2\bm{q}}$ are shorthand notation defined as:
\begin{align}
    &\epsilon_{\bm{k}_1\bm{k}_2\bm{q}} \equiv \epsilon_{\bm{k}_1+\bm{q}} + \epsilon_{\bm{k}_2-\bm{q}} - \epsilon_{\bm{k}_2} - \epsilon_{\bm{k}_1},
    \nonumber \\
    &O_{\bm{k}_1\bm{k}_2\bm{q}} \equiv c^\dagger_{\bm{k}_1+\bm{q}\uparrow}c^\dagger_{\bm{k}_2-\bm{q}\downarrow}c_{\bm{k}_2\downarrow}c_{\bm{k}_1\uparrow}.
\end{align}
Furthermore, the {\footnotesize$\sideset{}{'}\sum_{\bm{k}_1\bm{k}_2\bm{q}}$} represents summing over the non-divergent terms, that is, with non-zero denominator.
\end{theorem}
\vspace{0.5cm}

We noticed that the summands in Eq.~(\ref{eq:main theorem formula}) initially included terms that diverge when the denominator vanishes. These terms correspond to disconnected diagrams in standard QFT calculations, and to be rigorous, we analyzed these terms and found that they appear similarly in both the numerator and denominator; hence they cancel out as expected. Details of the analysis are presented in the Appendix.~C. Therefore, the summation {\footnotesize$\sideset{}{'}\sum_{\bm{k}_1\bm{k}_2\bm{q}}$} only sums over the non-divergent terms.

The most critical ramification of the operator-expanded S-matrix presented in Eq.~(\ref{eq:main theorem formula}) is the facilitation of an order-by-order perturbative approximation using quantum computers. 

We highlight the fact that the S-matrix is, in fact, an orthogonal matrix. This observation is rooted in the reality of both the coefficients and the creation and annihilation operators $c^\dagger_{\bm{k}\sigma}$ and $c_{\bm{k}\sigma}$ within Eq.~(\ref{eq:main theorem formula}), under the Jordan-Wigner transformation. Consequently, we postulate that the S-matrix can be successively approximated by a series of products of orthogonal matrices, which are inherently derived from the Hubbard interaction. These real orthogonal matrices can be conveniently constructed as follows:
\begin{equation}
\label{eq:ansatz unitary}
    U_{\bm{k}_1\bm{k}_2\bm{q}}(\theta) = e^{\theta \left(O_{\bm{k}_1\bm{k}_2\bm{q}} - O^\dagger_{\bm{k}_1\bm{k}_2\bm{q}}\right)}. 
\end{equation}
Verification that $U_{\bm{k}_1\bm{k}_2\bm{q}}(\theta)$  is an orthogonal matrix is straightforward, given that the generator is anti-Hermitian and the operators $O_{\bm{k}_1\bm{k}_2\bm{q}}$ and $O^\dagger_{\bm{k}_1\bm{k}_2\bm{q}}$ are real. Moreover, as detailed in Appendix.~B, we expand the exponential in Eq.~(\ref{eq:ansatz unitary}) into a Taylor series to achieve its closed form:
\begin{align}       
\label{eq:close form ansatz unitary}
U_{\bm{k}_1\bm{k}_2\bm{q}}(\theta) &=\mathbbm{1} + \sin\theta\biggl(O_{\bm{k}_1\bm{k}_2\bm{q}} - O^\dagger_{\bm{k}_1\bm{k}_2\bm{q}} \biggr)  
\\
&+(\cos\theta-1)\biggl(O_{\bm{k}_1\bm{k}_2\bm{q}}O^\dagger_{\bm{k}_1\bm{k}_2\bm{q}} + O^\dagger_{\bm{k}_1\bm{k}_2\bm{q}}O_{\bm{k}_1\bm{k}_2\bm{q}} \biggr).   \nonumber    
\end{align} 
Linking this closed form with Eq.~(\ref{eq:main theorem formula}), the first-order perturbation term in the latter can be reformulated into an anti-Hermitian expression (as discussed in Appendix.~D):
\begin{equation}
    -g\sideset{}{'}\sum_{\bm{k}'_1\bm{k}'_2\bm{q}'}\frac{1}{\epsilon_{\bm{k}'_1\bm{k}'_2\bm{q}'}}\biggl(O_{\bm{k}'_1\bm{k}'_2\bm{q}'}-O^\dagger_{\bm{k}'_1\bm{k}'_2\bm{q}'}\biggr).
\end{equation}
Analytical recovery of the first-order perturbation is achieved by applying a sequence of unitaries involving terms in the summation {\footnotesize $\sideset{}{'}\sum_{\bm{k}_1\bm{k}_2\bm{q}}$}, where the parameters of the unitaries satisfy:
\begin{equation}
\sin\theta_{\bm{k}'_1\bm{k}'_2\bm{q}'}=-\frac{g}{\epsilon_{\bm{k}'_1\bm{k}'_2\bm{q}'}}.
\end{equation}
The complexity escalates with the second and higher-order perturbation expansions, primarily due to the coupling of different indices in the coefficients at higher orders:
\begin{equation}
    \frac{1}{\sum_{j=1}^{m}\epsilon_{\bm{k}^{(j)}_1\bm{k}^{(j)}_2\bm{q}^{(j)}}}.
\end{equation}
The functional form of the higher-order expansions can still be approximated by sequential applications of the unitary operation as expressed in Appendix.~B. This iterative process allows the introduction of additional degrees of freedom, facilitating higher-order corrections without altering the lower-order perturbative results. For an in-depth examination of this technique, we direct the reader to Appendix.~D. Although there is no explicit analytical method to determine parameter assignments that recover all higher-order perturbations, ADAPT-VQE emerges as an exemplary numerical tool to discover the optimal parameters within these complex expansions. Our numerical experiments provide compelling evidence that ADAPT-VQE, especially when applied to the 
$k$-space representation of the Hubbard Hamiltonian, can systematically reconstruct the perturbation series order by order.

In the subsequent section, we present the comprehensive algorithm of ADAPT-VQE. This algorithm iteratively builds upon the ansatz, selecting the most significant operators to refine the approximation of the system's ground state. ADAPT-VQE's dynamic approach not only circumvents the need for a predetermined ansatz structure but also enhances the efficacy and efficiency of the simulation, which is particularly beneficial for complex quantum systems where traditional methods fall short.

\subsection{Variational Interaction-Picture S-matrix Ansatz}
In the section, we propose an algorithm that can perturbatively approximate the results in the Gell-mann and Low theorem. For graphical illustration of the algorithm, please refer to Fig. \ref{fig: flow chart}. This algorithm build on the theory we establish in the Appendix.~C, in which we explicitly write down the fermionic expansion of the normalized S-matrix.
\vspace{0.25cm}
\\
\noindent \textbf{1. Operator pool construction.}
In line with our previous discussions, the operators selected for inclusion in our pool are defined as follows:
\begin{align}
    &A_{\bm{k}_1,\bm{k}_2,\bm{q}}\equiv O_{\bm{k}_1\bm{k}_2\bm{q}}-O^\dagger_{\bm{k}_1\bm{k}_2\bm{q}}, 
    \nonumber\\
    &\forall \bm{k}_1\bm{k}_2\bm{q}, \hspace{0.25cm}\text{with } \epsilon_{\bm{k}_1\bm{k}_2\bm{q}}\neq0.
\end{align}
As detailed in Appendix.~B, the corresponding unitary operator is derived as:
\begin{align}       
U_{\bm{k}_1\bm{k}_2\bm{q}}(\theta) &= e^{\theta (O_{\bm{k}_1\bm{k}_2\bm{q}} - O^\dagger_{\bm{k}_1\bm{k}_2\bm{q}})} \nonumber \\ &=\mathbbm{1} + \sin\theta\biggl(O_{\bm{k}_1\bm{k}_2\bm{q}} - O^\dagger_{\bm{k}_1\bm{k}_2\bm{q}} \biggr).  
\\
&+(\cos\theta-1)\biggl(O_{\bm{k}_1\bm{k}_2\bm{q}}O^\dagger_{\bm{k}_1\bm{k}_2\bm{q}} + O^\dagger_{\bm{k}_1\bm{k}_2\bm{q}}O_{\bm{k}_1\bm{k}_2\bm{q}} \biggr).   \nonumber    
\end{align} 
The aim is to utilize these unitary operators to accurately approximate the normalized time evolution operator, $U_{\mathcal{I}}(0,-\infty)$. A significant distinction between our method and the original ADAPT-VQE algorithm is the exploitation of translational invariance in $k$-space, which effectively reduces the complexity of four-fermion operators to three indices. This reduction yields a more manageable pool size scaling as $\mathcal{O}(n^3)$ pool size. Furthermore, our theoretical findings suggest that the normalized time evolution operator consists solely of the Hubbard terms $O_{\bm{k}_1\bm{k}_2\bm{q}}$, thus eliminating the necessity for two-fermion operators. In contrast, the original ADAPT-VQE is predicated on Unitary Coupled Cluster theory, which necessitates a more extensive pool of $\mathcal{O}(n^4)$ operators due to the inclusion of both single and double excitation operators.
\vspace{0.25cm}
\\
\noindent \textbf{2. Prepare non-interacting ground state.} To prepare the non-interacting ground state, it is necessary to diagonalize the hopping terms (formulated in real space). This can be accomplished by Fourier transforming the hopping terms into k-space:
$$
\text{FT}^\dagger \biggl(-t\sum_{\langle i,j\rangle, \sigma} a^\dagger_{i\sigma}a_{j\sigma}+a^\dagger_{j\sigma}a_{i\sigma}\biggr) \text{FT}=\sum_{\bm{k}\sigma}\epsilon_{\bm{k}}c^\dagger_{\bm{k}\sigma}c_{\bm{k}\sigma}.
$$
By arranging the dispersion relation $\epsilon_{\bm{k}}$ in ascending order, we can prepare the non-interacting ground state by applying Pauli-$X$ operators to the first $N_{\text{up}}$ and $N_{\text{down}}$ orbitals. A detailed discussion on this is available in Appendix.
\vspace{0.25cm}
\\
\noindent\textbf{3. Gradient evaluation.} In general, the gradient of the unitary operator, initialized with $\theta_{\bm{k}_1\bm{k}_2\bm{q}}=0$, can be evaluated as follows:
\begin{equation}
    g_{\bm{k}_1\bm{k}_2\bm{q}}=\frac{\partial \langle H \rangle_{\bm{\theta}}}{\partial \theta_{\bm{k}_1\bm{k}_2\bm{q}}}=\langle \psi(\bm{\theta})|[H, A_{\bm{k}_1\bm{k}_2\bm{q}}]|\psi(\bm{\theta})\rangle,
\end{equation}
where $\bm{\theta}$ represents the parameters of the operators selected in the previous epoch, denoted as $\bm{\theta}=(\theta^{(1)},\cdots,\theta^{(\ell)})$. After evaluating the gradients over the operators in the pool, we check if the maximum gradient falls below the threshold $\epsilon_1$. If this condition is satisfied, we consider our algorithm converged, and thus, we terminate the entire algorithm.
\vspace{0.25cm}
\\
\noindent\textbf{4. Select Operators.} Once we have computed the gradients for all operators in the pool, we select the operators with gradient values exceeding $r$ times the maximum gradient, i.e.,
\begin{equation}
    S=\Biggl\{A_{\bm{k}_1\bm{k}_2\bm{q}} \Biggr| \hspace{0.1cm} |g_{\bm{k}_1\bm{k}_2\bm{q}}| \geq r\cdot\max_{\bm{k}_1\bm{k}_2\bm{q}} |g_{\bm{k}_1\bm{k}_2\bm{q}}|\Biggr\}.
\end{equation}
In our implementation, we have chosen to set the parameter $r$ to a value of 0.1. The rationale behind this choice stems from the observation that, initially, there exist several operators with gradients of equal magnitude, while other operators exhibit zero gradients. This particular setting is effective in ensuring the selection of all operators with non-zero gradients in the first epoch.
\vspace{0.25cm}
\\
\noindent\textbf{5. Append new operators.} Once we have selected the operators, we append the new unitary operations to our previous circuit. It's worth noting that our wave function is in k-space, but for the sake of reducing measurement complexity, we apply the Fourier transform to real space. Therefore, the appended unitary should be inserted between the previous unitary and the Fourier transform circuit.
\vspace{0.25cm}
\\
\noindent\textbf{6. Optimization.} We re-optimize the entire ansatz, with the parameters of the appended unitary initialized at 0 to avoid perturbing the previous state. Instead of optimizing the circuit in a single iteration and proceeding to step 2, we optimize our circuit until it converges. This step is crucial for the success of our numerical experiments and ensures convergence.

\section{Experiment Results}
In this section, we present the details of our numerical experiment and demonstrate how our method effectively finds the ground state and reproduces the effects predicted by the Gell-Mann \& Low theorem.

\subsection{2D Fermi-Hubbard Model}
The Hamiltonian of the 2D Fermi-Hubbard model is described by the following equation:
\begin{equation}
\label{eq: hubbard model real space}
    H_U = -t\sum_{\langle i,j \rangle, \sigma}(a^\dagger_{i\sigma}a_{j\sigma} + a^\dagger_{j\sigma}a_{i\sigma}) + U\sum_{i}n_{i\uparrow}n_{i\downarrow}.
\end{equation}
In this study, we focus specifically on the parameters $t = 1$ and $U = 2, 4, 6$. The operators $a^\dagger_{i\sigma}$ and $a_{i\sigma}$ represent creation and annihilation at site $i$ with spin $\sigma \in \{\uparrow, \downarrow\}$, respectively, while $n_{i\sigma}$ denotes the number operator. The first term in the equation represents the nearest-neighbor hopping interaction, and the second term signifies the Coulomb repulsion when two electrons occupy the same site. Our study primarily focuses on the $k$-space representation, meaning that all unitaries and basis states are expressed in the context of $k$-space. This approach may lead to $\mathcal{O}(n^3)$ measurement complexity. To reduce the measurement overhead, we perform a Fourier transform to real space in the final step of the quantum circuit, allowing us to measure the Hamiltonian in Eq.~(\ref{eq: hubbard model real space}) with a reduced measurement complexity of $\mathcal{O}(n)$. We examine grid sizes of $2 \times 2$, $2 \times 3$, $2 \times 4$, and $3 \times 3$. For $N_x \times N_y$ cases, we impose periodic boundary conditions when either $N_x$ or $N_y$ is greater than 2, and open boundary conditions when $N_x$ or $N_y$ is equal to 2.

\begin{figure}
    \centering
    
    \subfigure[]{
        \includegraphics[width=0.4\linewidth]{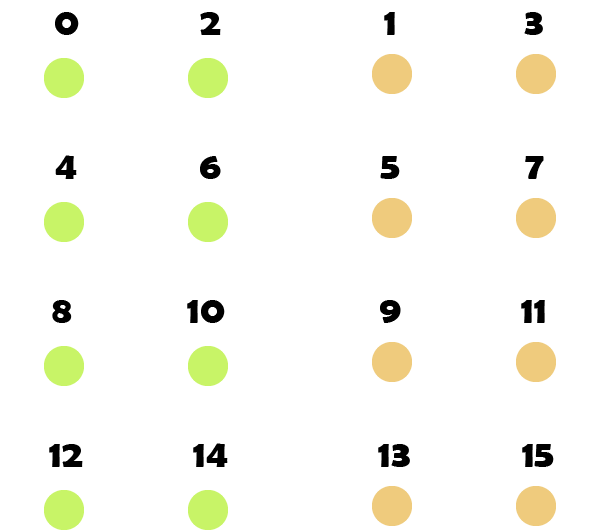}
        \label{fig:OF convention 1}
    }
    \hspace{0.5cm}
    \subfigure[]{
        \includegraphics[width=0.45\linewidth]{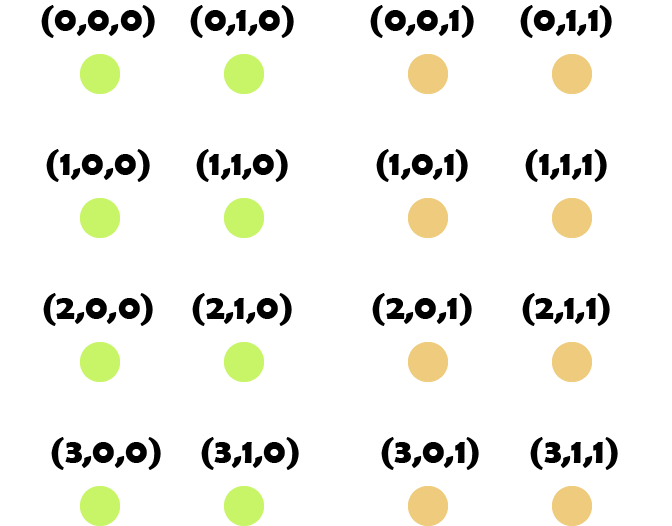}
        \label{fig:OF convention 2}
    }
    \caption{Grid index conventions. (a) Shows the indexing convention used by OpenFermion for mapping spin orbitals to qubits. (b) Displays the tuples representing the real space or \(k\)-space coordinates along with the spin, where the first two elements are the coordinates and the last one denotes the spin state.}
    \label{fig:OF convention}
\end{figure}

\begin{figure}[t]
    \centering
    \includegraphics[width=0.45\textwidth]{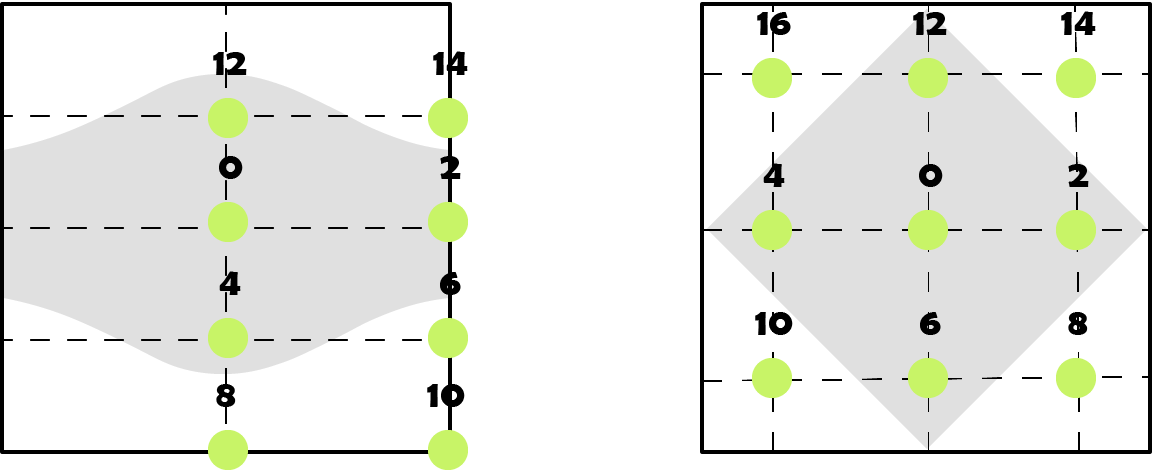}
    \caption{Brillouin zone of the non-interacting Fermi-Hubbard Hamiltonian, with only the up-spin orbitals shown. (a) Brillouin zone with the $x$-axis under open boundary conditions and the $y$-axis under periodic boundary conditions, where the Fermi surface is defined by the condition \(\cos k_x + 2\cos k_y = 0\). (b) Brillouin zone with both axes subject to either periodic or open boundary conditions, with the Fermi surface given by \(\cos k_x+\cos k_y=0\). 
    } 
    \label{fig:BZ}
\end{figure}

To represent the fermionic system on the quantum computer, here we use Jordan-Wigner transform, in which each spin orbital corresponds to a qubit in the 1D line. The hopping and Hubbard terms in $H_U$ under the Jordan-Wigner transformation are respectively: 
\begin{align}
&a^\dagger_{i\sigma}a_{j\sigma}+a^\dagger_{j\sigma}a_{i\sigma} \mapsto \frac{X_{i\sigma}X_{j\sigma}+Y_{i\sigma}Y_{j\sigma}}{2}\prod_{i<k<j}Z_{k\sigma}, 
\\
&n_{i\uparrow}n_{i\downarrow}\mapsto\frac{1}{4}(I+Z_{i\uparrow}+Z_{i\downarrow}+Z_{i\uparrow}Z_{i\downarrow}).
\end{align}
For the ansatz operators, the corresponding transform in the $k$-space is:
\begin{footnotesize}
\begin{align}
    &O_{\bm{k}_1\bm{k}_2\bm{q}}-O^\dagger_{\bm{k}_1\bm{k}_2\bm{q}} \mapsto
    \nonumber \\
    &\frac{1}{8}\biggl(X_{\bm{k}_1+\bm{q}\uparrow} Y_{\bm{k}_2-\bm{q}\downarrow} X_{\bm{k}_2\downarrow} X_{\bm{k}_1\uparrow}+Y_{\bm{k}_1+\bm{q}\uparrow} X_{\bm{k}_2-\bm{q}\downarrow} X_{\bm{k}_2\downarrow} X_{\bm{k}_1\uparrow}
    \nonumber \\
    &\hspace{0.2cm}+Y_{\bm{k}_1+\bm{q}\uparrow} Y_{\bm{k}_2-\bm{q}\downarrow} Y_{\bm{k}_2\downarrow} X_{\bm{k}_1\uparrow}+Y_{\bm{k}_1+\bm{q}\uparrow} Y_{\bm{k}_2-\bm{q}\downarrow} X_{\bm{k}_2\downarrow} Y_{\bm{k}_1\uparrow}
    \nonumber \\
    &\hspace{0.2cm}-X_{\bm{k}_1+\bm{q}\uparrow} X_{\bm{k}_2-\bm{q}\downarrow} Y_{\bm{k}_2\downarrow} X_{\bm{k}_1\uparrow}-X_{\bm{k}_1+\bm{q}\uparrow} X_{\bm{k}_2-\bm{q}\downarrow} X_{\bm{k}_2\downarrow} Y_{\bm{k}_1\uparrow}
    \nonumber \\
    &\hspace{0.2cm}-Y_{\bm{k}_1+\bm{q}\uparrow} X_{\bm{k}_2-\bm{q}\downarrow} Y_{\bm{k}_2\downarrow} Y_{\bm{k}_1\uparrow}-X_{\bm{k}_1+\bm{q}\uparrow} Y_{\bm{k}_2-\bm{q}\downarrow} Y_{\bm{k}_2\downarrow} Y_{\bm{k}_1\uparrow}
    \biggr)
    \nonumber \\
    &\prod_{\bm{k} \in (\bm{k}_1,\bm{k}_1+\bm{q})} Z_{\bm{k}\uparrow} \prod_{\bm{k}' \in (\bm{k}_2, \bm{k}_2-\bm{q})} Z_{\bm{k}'\downarrow}. 
\end{align}
\end{footnotesize}
\\
It is important to note that the eight Pauli strings commute with each other. Therefore, the unitary $U_{\bm{k}_1\bm{k}_2\bm{q}}(\theta)$ can be implemented without Trotter error. The labeling of the spin orbitals follows the convention used in OpenFermion, as presented in Fig. \ref{fig:OF convention}.

\subsection{Non-Interacting Ground State}
We investigate the non-interacting ground state of the Fermi-Hubbard Hamiltonian at half-filling, with a total of \( N \) electrons distributed across \( N \) sites. For scenarios where the \( x \)-axis is subject to open boundary conditions while the \( y \)-axis has periodic boundary conditions (specifically for cases \( 2 \times 3 \) and \( 2 \times 4 \)), the Fourier transform of the hopping terms is given by:
\begin{equation}
    H_0 = -t\sum_{\bm{k}, \sigma}\left(\cos k_x + 2\cos k_y \right) c^\dagger_{\bm{k}\sigma}c_{\bm{k}\sigma}.
\end{equation}
For \( N_y = 3 \), the Fermi surface exhibits double degeneracy, leading to a ground state degeneracy of 4. Conversely, for \( N_y = 4 \), the non-interacting ground state is uniquely defined.

When both axes are subject to open or periodic boundary conditions (\( 2 \times 2 \) and \( 3 \times 3 \)), the Fourier transform of the hopping term is:
\begin{equation}
    H_0 = -2t\sum_{\bm{k}, \sigma}\left(\cos k_x + \cos k_y \right) c^\dagger_{\bm{k}\sigma}c_{\bm{k}\sigma}.
\end{equation}
In the \( 2 \times 2 \) case, the Fermi surface is doubly degenerate, resulting in a total ground state degeneracy of 4. For the \( 3 \times 3 \) configuration, we select 5 up-spin electrons and 4 down-spin electrons, which results in a total degeneracy of 4. Note that for each degenerate non-interacting ground state, the initial occupation is chosen according to the index order of the degenerate spin orbitals on the Fermi surface.

\begin{figure*}
    \centering
    
    \subfigure[Hubbard model 2x4 (t=1, U=2)]{
        \includegraphics[width=0.475\linewidth]{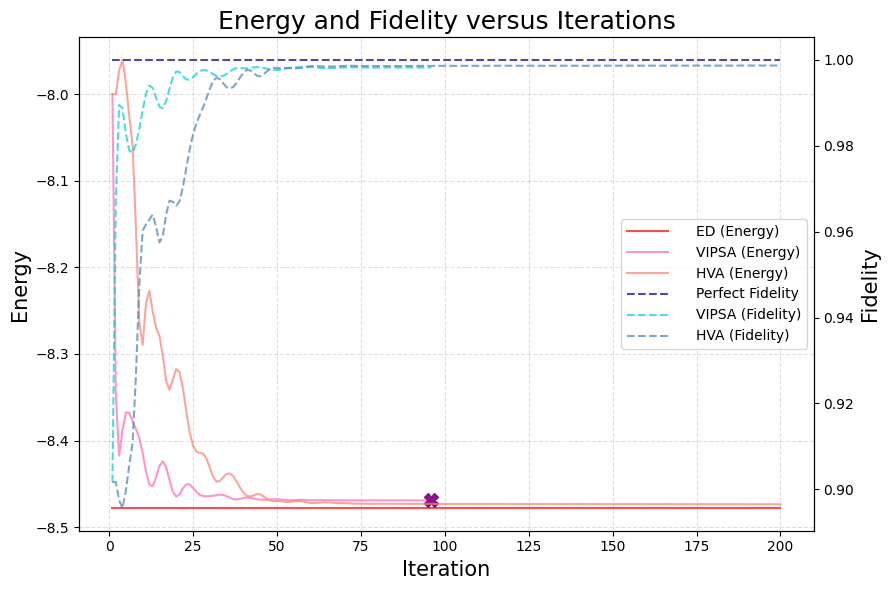}
        \label{fig:2x4 U=2}
    }
    \subfigure[Hubbard model 2x4 (t=1, U=4)]{
        \includegraphics[width=0.475\linewidth]{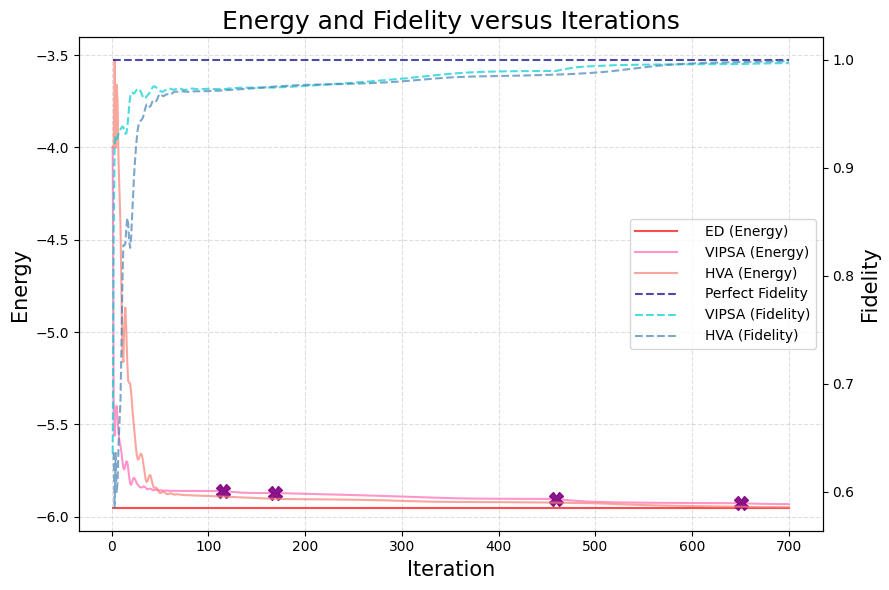}
        \label{fig:2x4 U=4}
    }
    \subfigure[Hubbard model 2x4 (t=1, U=6)]{
        \includegraphics[width=0.475\linewidth]{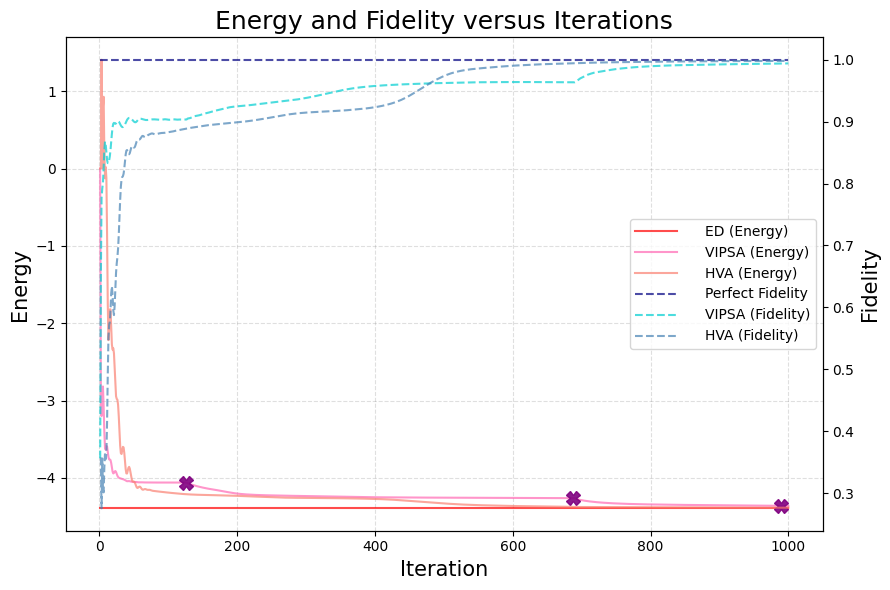}
        \label{fig:2x4 U=6}
    }
    \subfigure[Hubbard model 3x3 (t=1, U=2)]{
        \includegraphics[width=0.475\linewidth]{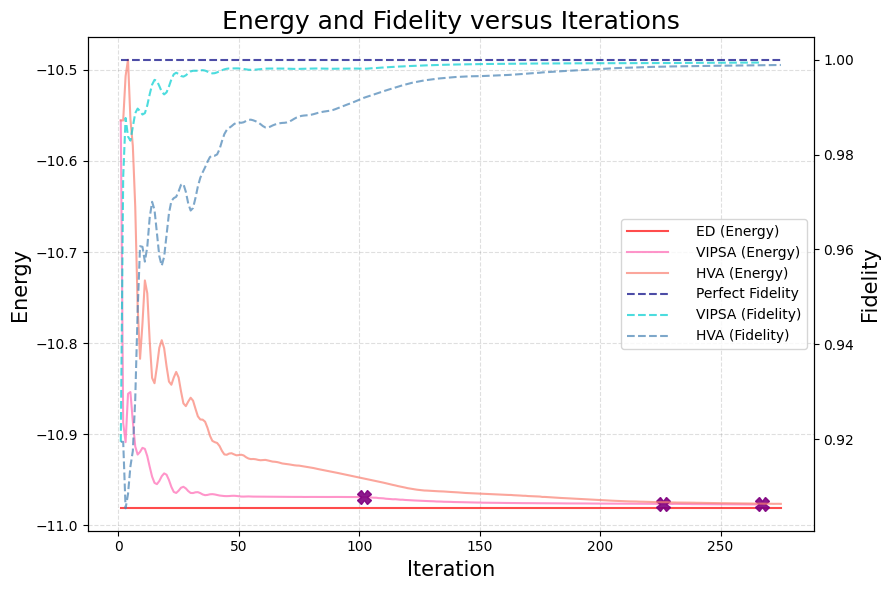}
        \label{fig:3x3 U=2}
    }
    \subfigure[Hubbard model 3x3 (t=1, U=4)]{
        \includegraphics[width=0.475\linewidth]{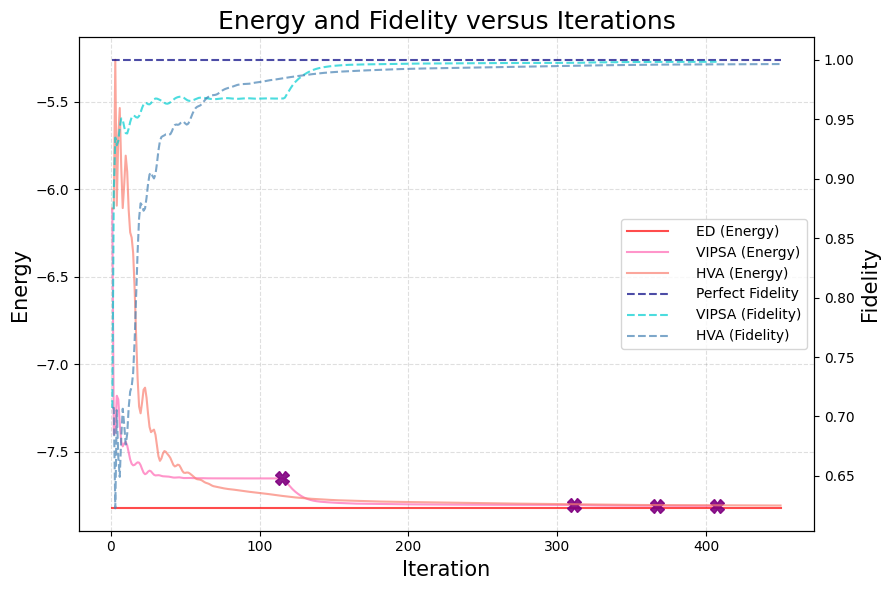}
        \label{fig:3x3 U=4}
    }
    \subfigure[Hubbard model 3x3 (t=1, U=6)]{
        \includegraphics[width=0.475\linewidth]{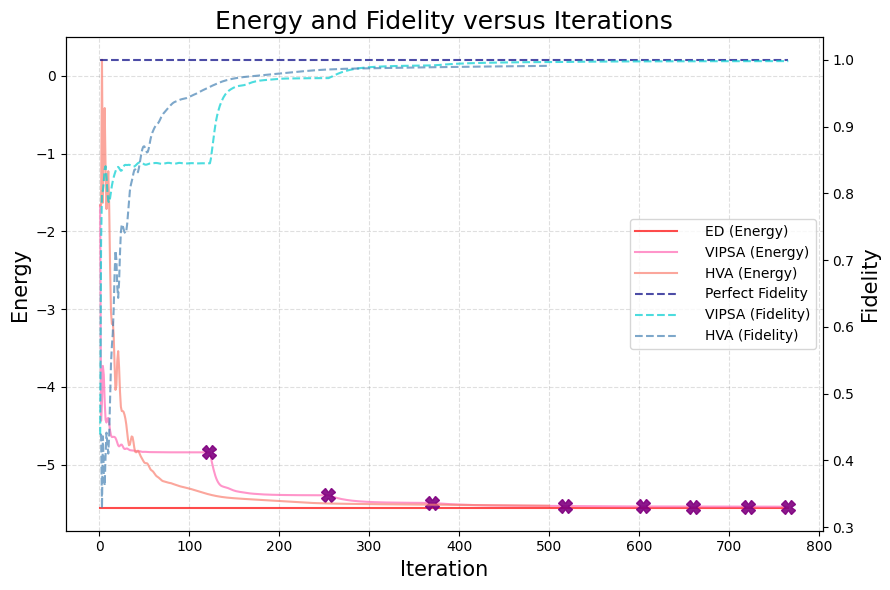}
        \label{fig:3x3 U=6}
    }
    \caption{Simulation results comparing the Variational Interaction Picture S-matrix Ansatz (VIPSA) with the Hamiltonian Variational Ansatz (HVA) for the Hubbard model on $2\times4$ and $3\times3$ grids with Hubbard interaction strengths $U=2, 4, 6$. The left $y$-axis indicates the energy while the right $y$-axis labels the fidelity of the state. The energies obtained by VIPSA, HVA, and exact diagonalization (ED) are shown in pink, orange, and red lines respectively.Purple cross marks signify the converged energies at each epoch. Fidelities for VIPSA and HVA are indicated by light blue and sky blue lines, with the perfect fidelity benchmark in dark blue. Results for the $2\times2$ and $2\times3$ grid sizes are provided in the Appendix.~G It is noted that the exact ground states for the $3\times3$ Hubbard model have a 4-fold degeneracy. Therefore, the fidelity on this grid size is calculated by summing the fidelities of the four ground states: $|\langle\psi|\Psi_1\rangle|^2+|\langle\psi|\Psi_2\rangle|^2+|\langle\psi|\Psi_3\rangle|^2+|\langle\psi|\Psi_4\rangle|^2$.}
    \label{fig: simulation result}
\end{figure*}

\subsection{Numerical Results}
In this section, we present our numerical results and benchmark our method against the Hamiltonian Variational Ansatz (HVA), the current state-of-the-art approach for simulating the Fermi-Hubbard model. We employ the ADAM optimizer, favored in the deep learning field, with learning rate $10^{-2}$ for optimizing our ansatz. All cases considered are under the premise that exact energy can be computed in each iteration, equivalent to having an effectively infinite number of measurements. For our method, we set the epoch convergence tolerance ($\epsilon_2$) to $10^{-2}$, and the overall terminal tolerance ($\epsilon_1$) to $10^{-2}$. In our implementation of the HVA for various grid configurations, we standardize the number of layers ($S$) at 10. The number of parameters in the HVA is contingent on the grid size. Specifically, for $2\times N_y$ grids, a single parameter suffices for exponentiating the horizontal hopping terms. For even $N_y$ values greater than 2, two parameters are necessary to exponentiate the vertical hopping terms. Conversely, for odd $N_y$ values greater than 1, three parameters are required for this purpose. 

The simulation outcomes for the Hubbard model on grid sizes of $2\times4$ and $3\times3$, with interaction strengths $U=2, 4, 6$, are depicted in Fig.~\ref{fig: simulation result}. Results for the $2\times2$ and $2\times3$ configurations are detailed in Appendix.~G. In the figures, the energy of the model wavefunction is plotted on the left y-axis, and the fidelity is represented on the right y-axis. Energy curves are illustrated with a red solid line, while fidelity is traced with a dashed blue line. Each purple cross symbol marks the converged energy for the VIPSA at respective epochs, providing a clear visual indicator of the optimization progress at each step.

We compare the energy and fidelity convergence in the weak and strong interaction region. In the domain of weak interaction, our model generally exhibits faster convergence compared to HVA. This phenomenon may be attributable to our ansatz's ability to capture the first-order perturbation expansion initially. When interaction strength is modest, first-order perturbation typically offers a highly accurate approximation. Our conjecture that our model captures the first-order perturbation theory is supported by the numerical finding that, initially, only a few operators contribute significantly to the gradient, with the remaining operators showing a zero gradient. During the first epoch, we observed that the number operators and the spin flip operators—which cause divergent terms in the S-matrix are excluded from the operator selection process. Moreover, excitation operators that annihilate the non-interacting ground state are also omitted. This finding aligns with our theoretical expectations, reinforcing our belief that the first epoch effectively recovers the first-order perturbation.

In regions of stronger interaction, our model demonstrates an exceptional convergence rate in the initial epochs, suggesting that our approach, along with the theoretical underpinnings outlined in Appendix.~C, may hold the potential to surpass the QAOA-like quantum algorithm. While the convergence rate of our model is slightly slower than that of the Hamiltonian Variational Ansatz (HVA) in later epochs, this difference may not pose a significant concern. Adjustments to the convergence criteria could allow our model to select operators earlier in the process. The observed decrease in the convergence rate during the later epochs can be linked to the limitations of first-order perturbation theory as an approximation method, especially under conditions of heightened interaction strengths. This is particularly noticeable in Fig.~\ref{fig:2x4 U=6}, \ref{fig:3x3 U=4}, and~\ref{fig:3x3 U=6}, where there are discernible plateaus (indicated by the pink line) in the initial epoch. An additional insight from our numerical results is that a higher interaction strength necessitates more epochs (a greater number of selections). We associate this with the slower decay of higher-order perturbations as the interaction strength intensifies. Conversely, the increased number of iterations required by HVA to converge to the ground state may also signal the onset of strong interactions.

It is critical to recognize that the proper preparation of the initial state, with the correct spin symmetry, is essential for the convergence of both models to the ground state. This requirement is particularly evident in the case of the $2\times3$ grid, where the true ground state is a spin triplet state (with $S^2=2$). If the initial state is prepared with spin symmetry, specifically with an equal number of up-spin and down-spin electrons, the model fails to converge to the ground state. This limitation arises because the horizontal and vertical hopping terms, as well as the Hubbard terms, all commute with $S_z$ and $S^2$ (detailed proof can be found in Appendix.~F). Consequently, for the HVA, the model cannot transition out of the subspace once the initial spin symmetry is set, underscoring the necessity of setting the correct initial conditions for successful simulations. In contrast, for the VIPSA, the generator $O_{\bm{k}_1\bm{k}_2\bm{q}}-O^\dagger_{\bm{k}_1\bm{k}_2\bm{q}}$ commutes with $S_z$ but not with $S^2$. Our simulation results exhibit some symmetry-breaking effects throughout the optimization process. Nevertheless, initiating the process with the correct spin symmetry is still essential, as the symmetry-breaking impact on $S^2$ is relatively weak, making the evolution from a spin symmetric state to a spin triplet state challenging. 

Another critical aspect to highlight is that HVA exhibits a zero gradient at the non-interacting ground state initially. This phenomenon has been explored in existing literature, which shows that exponentiating a real operator, when the initial state is a real wave function, will lead to a zero initial gradient. For a detailed proof of this property, the reader is directed to the Appendix.~E. This behavior is evident in the simulation results shown in Fig. \ref{fig: simulation result}, where the energy of HVA peculiarly increases during the initial epochs. This counterintuitive increase in energy is ascribed to the workings of the ADAM optimizer, which, in the face of an initial zero gradient, adaptively adjusts the step size, inadvertently assigning a disproportionately large initial step to HVA. As HVA exits the manifold of real wave functions, the oversized step causes the energy to surge in the first few epochs. Conversely, our model demonstrates a more robust and stable training trajectory, with rapid convergence observed from the onset. Furthermore, our findings offer a crucial insight into the trajectory from the non-interacting to the interacting eigenstate. Commonly, it is presumed that the Trotter expansion of the adiabatic time evolution offers the quickest path. However, our theoretical analysis reveals that the S-matrix, which connects the interacting and non-interacting eigenstates, is surprisingly a real and orthogonal matrix. The ansatz form of HVA, as expressed in the equation:
\begin{equation}
    \Psi_T=\prod_{b=1}^{S}\biggl(U_{U} \left(\frac{\theta^b_U}{2}\right)U_h\left(\theta^b_h\right)U_v\left(\theta^b_v\right)U_{U} \left(\frac{\theta^b_U}{2}\right)\biggr)\Psi_I,
\end{equation}
is inherently a unitary matrix. This implies that HVA may not be the most efficient ansatz for reaching the ground state.
\\
\section{Discussion \& Conclusion}
This study marks a significant advancement in quantum computing and simulations, underpinned by the derivation of an analytic form for the normalized time evolution operator within the framework of the Gell-Mann \& Low theorem. This allowed us to systematically handle the divergent terms in the time evolution operator, and rigorously show that the divergent terms correspond to disconnected vacuum diagrams and cancel out in an similar manner to traditional QFT methodology. The culmination of this is a well-behaved, non-divergent, operator-expansion of the S-matrix; thereby providing a path for reconstruction of the Dyson series on a quantum computer. 

Our further theoretical analysis demonstrates that by applying a sequence of ansatz unitaries $U_{\bm{k}_1\bm{k}_2\bm{q}}(\theta)$, the entire functional form of the Dyson series can be recovered order by order, including the analytical form for the first-order perturbation. However, the exact expressions for parameter assignment in higher-order perturbation terms remains a challenge for future work. 

Another significant contribution of our research is the introduction of the Variational Interaction Picture S-matrix Ansatz (VIPSA), inspired by the ADAPT-VQE algorithm. This algorithm integrates perturbation theory into the core of the variational quantum eigensolver paradigm, a methodology largely unexplored in the realm of quantum computing. By integrating perturbation theory into quantum simulation processes, we have ventured into new territory, enhancing the exploration of complex quantum systems with greater efficiency and precision. Empirical validation of VIPSA, particularly with the 2D Fermi-Hubbard model, confirms its practicality. The algorithm is adept at determining the exact ground state in strongly correlated systems. Upon examining the operators chosen in the first epoch, we find that these operators can indeed reconstruct the first-order perturbation expansion. Furthermore, these operators exclude the divergent terms found in the time evolution operator, and any operator that would annihilate the initial state is also excluded, consistent with our theoretical predictions.

Our numerical investigations reveal that our method consistently achieves robust initial convergence rates, outpacing the Hamiltonian Variational Ansatz (HVA). Theoretical analysis discloses a zero initial gradient for HVA, causing fluctuating energy levels during initial optimization. This pattern intimates that strategies similar to the Quantum Approximate Optimization Algorithm (QAOA) might not adopt the most expedient route to the ground state. Based upon the key insight that the Hubbard term is a real symmetric matrix, and that the groundstate wavefunction is real in the JW basis, we conjecture that the orthogonal form of the S-matrix that we proposed gives a more optimal convergence in comparison to the standard unitary operators produced via conventional Trotterization process in QAOA and HVA.

The insights garnered from our theoretical and empirical work suggest a novel blueprint for quantum algorithm development, pioneering the integration of perturbation theory into the algorithmic framework. It is plausible that our theoretical framework could extend to more complex interactions by substituting the uniform interaction term $\frac{U}{N}$ with a multifaceted interaction model $V_{\bm{k}_1\bm{k}_2\bm{q}}$. Particularly, as system intricacies amplify with the incorporation of extended-range hoppings or beyond on-site Hubbard interactions, our method may exhibit a significant performance edge, given that such terms seamlessly meld into the interaction term in the interaction picture, without compounding the complexity of the operator pool.

Moreover, our algorithm could emerge as a viable alternative to QAOA, especially since current theories do not provide assurances about the evolution time required by QAOA. Our theoretical framework could offer a novel means to estimate the resources and overall overhead necessary for quantum computations, potentially steering the field towards more efficient quantum algorithmic designs. In sum, the outcomes of this investigation are manifold. We have not only forged a new and novel operator formulation of the normalized time evolution operator, but have also unveiled a trailblazing methodological application for quantum computing simulations. The implications of these advancements are poised to have extensive impact across quantum field theory, condensed matter physics, and the expansive quantum computing landscape, laying a robust foundation for future scholarly pursuits and technological advancements.

\bibliographystyle{apsrev4-2}
\bibliography{references.bib}
\end{document}